\documentclass[twocolumn,aps]{revtex4}
\usepackage{amssymb,graphicx}
\begin{document}

\title{Extend Special Relativity to the Superluminal Case}
\author{Z. C. Tu} \email[Now in
Department of Physics, Beijing Normal University.\,\,\,\, Email
address: ]{ tuzc@bnu.edu.cn} \affiliation{Institute of Theoretical
Physics, The Chinese Academy of Science, P.O.Box 2735 Beijing
100080, China}
 \author{Z. Y. Wan}
\address{Faculty 404, Beijing University of Aeronautics and Astronautics, Beijing, 100083, P. R. China}

\begin{abstract}
First, we extend the special relativity into the superluminal case
and put forward a superluminal theory of kinematics, in which we
show that the temporal coordinate need exchanging with one of the
spatial coordinates in a superluminal inertial frame, and that the
coordinate transformations from any superluminal inertial frame to
the rest frame (here rest just says in a relative sense) are the
same as the Lorentz transformations from some normal inertial frame
to the rest frame. Consequently, the causality can not be violated.
Secondly, we investigate the superluminal theory of dynamics and
find that the total energy of any object moving at a speed of $v$
(faster than the speed of light in vacuum $c$) is equal to the total
energy of that object moving at a speed of $u\,(u<c)$ provided that
the product of two speeds satisfy $uv=c^{2}$. Lastly, we conjecture
that this superluminal theory can give a novel interpretation to the
essence of matter waves put forward by de Broglie.
\end{abstract}
\maketitle

Since the special relativity (SR) was put forward by Einstein
\cite{a} in 1905, the speed of light in vacuum $c$ has always been
regarded as the limit of particle velocity. As is so often pointed
out, the causality will be violated if there are tachyons in our
world \cite{b} . In fact, nature has exhibited many superluminal
phenomena to us, especially in astronomy \cite{c,d} , but they do
not imply the genuine existence of superluminal objects.
Furthermore, there exist various proposals for observing
faster-than-$c$ propagation of light using anomalous dispersion near
an absorption line and linear gain lines, or tunnelling barriers
\cite{e,f,g,h} . It is notable that, using gain-assisted linear
anomalous dispersion, the group velocity of a laser pulse
propagation in atomic caesium gas can exceed $c$ and even become
negative, while the shape of the pulse is preserved \cite{i} . But
these proposals can not be explicitly interpreted as the motion of
tachyons. The same case also occurs in the propagation of localized
microwaves \cite{j} . All these seem to suggest that the speed of
any moving object cannot exceed $c$ indeed. But virtually, Einstein
just tells us that $c$\ is the limit of particle velocity in vacuum.
As we know, the limit may indicate the upper bound of the particle
velocity which corresponds to the SR, while it may mean the lower
bound. In our paper, we aim to develop a superluminal theory which
abides by causality and does not contradict most excellent
achievements of SR.

\begin{figure}[htp!]
\includegraphics[width=10cm]{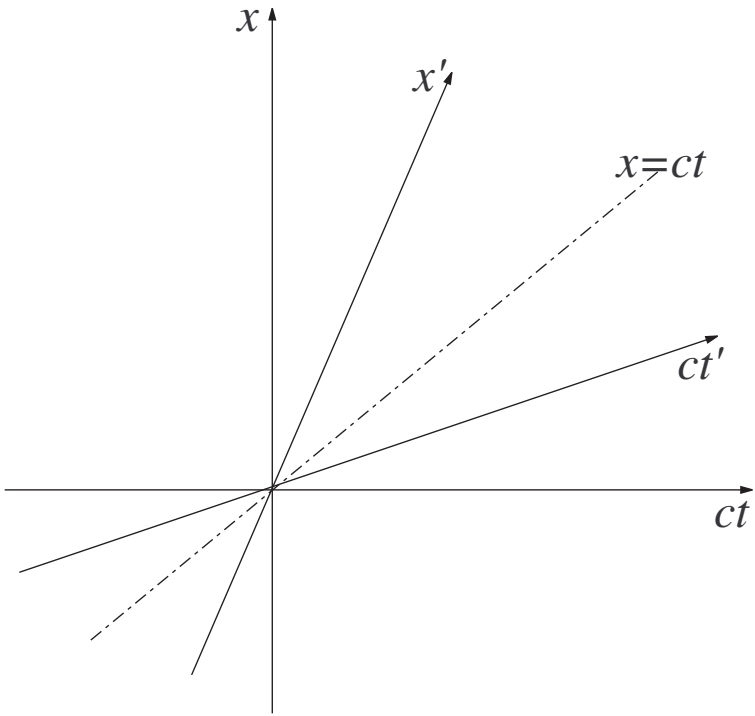}\caption{The geometrical
relation between the pairs $(x,ct)$ and $(x^{\prime },ct^{\prime
})$. \label{fig1}}
\end{figure}

Let $S$ denote the rest frame in which $x,y$ and $z$ are three spatial
coordinates and $t$ is the temporal coordinate. $S^{\prime }$ is an inertial
frame moving at the speed of $v$ along $x$-axis of $S$. Denote $x^{\prime
},y^{\prime },z^{\prime }$ and $t^{\prime }$ as three spatial coordinates
and temporal coordinate respectively in $S^{\prime }.$ At the initial time $%
t=t^{\prime }=0$, the origins of two frames are superposed each other. In
terms of SR, the Lorentz transformations between $S$ and $S^{\prime }$ can
be expressed as follow:

\begin{equation}
\left\{
\begin{array}{l}
x^{\prime }=\frac{1}{\sqrt{1-\left( v/c\right) ^{2}}}(x-\frac{v}{c}ct)\qquad
(\rm{i}) \\
ct^{\prime }=\frac{1}{\sqrt{1-\left( v/c\right) ^{2}}}(ct-\frac{v}{c}x)\quad
\ \ (\rm{ii}) \\
y^{\prime }=y\qquad \qquad \qquad (\rm{iii}) \\
z^{\prime }=z\qquad \qquad \qquad (\rm{iv})
\end{array}
\right. ,\label{eq1}
\end{equation}
where $c$ is the light speed in vacuum and $v<c$. Eq. (\ref{eq1}-iii) and Eq. (\ref{eq1}-iv)
tell us that we merely need to consider the relation between the pairs $%
(x,ct)$ and $(x^{\prime },ct^{\prime })$. Their geometrical relation is
shown in Fig.\ref{fig1}. The positions of $x^{\prime }$-axis and $ct^{\prime }$-axis
in the coordinate frame $(x,ct)$ are obtained:

\begin{equation}
\left\{
\begin{array}{l}
x^{\prime }-\rm{axis}:ct^{\prime }=0\Longrightarrow x=\frac{c}{v}ct \\
ct^{\prime }-\rm{axis}:x^{\prime }=0\Longrightarrow x=\frac{v}{c}ct
\end{array}
\right. .\label{eq2}
\end{equation}
Eq. (\ref{eq2}) is valid if $v<c$, from which we can deduce that, in Fig.\ref{fig1}, both $%
x^{\prime }$-axis and $ct^{\prime }$-axis are rotating to the line $(x=ct)$
with the increasing of $v$, and we can suppose that they superpose with the
line $(x=ct)$ if $v=c$. Moreover, we can imagine that, if $v>c$, $x^{\prime
} $-axis enters the domains between that line and $ct$-axis, and $ct^{\prime
}$-axis enters the other domains. In this case, we need to redefine the
concepts of $x^{\prime }$ and $ct^{\prime }$ in $S^{\prime }$.

We extrapolate Lorentz transformations to the superluminal case $(v>c)$:

\begin{equation}
\left\{
\begin{array}{l}
x^{\prime }=\frac{1}{i\sqrt{\left( v/c\right) ^{2}-1}}(x-\frac{v}{c}%
ct)\qquad (\rm{i}) \\
ct^{\prime }=\frac{1}{i\sqrt{\left( v/c\right) ^{2}-1}}(ct-\frac{v}{c}%
x)\quad \ \ (\rm{ii})
\end{array}
\right. ,\label{eq3}
\end{equation}
where $i$ $(i^{2}=-1)$ is the imaginary unit.

Consider an invariant of Minkowski space-time:

\begin{eqnarray}
\Delta s^{2}=x^{2}+(ict)^{2}=x^{\prime 2}+(ict^{\prime })^{2},\label{eq4}
\end{eqnarray}
where the terms $y^{2},z^{2},y^{\prime 2},z^{\prime 2}$ are omitted
for the sake of Eq. (\ref{eq1}-iii) and Eq. (\ref{eq1}-iv). Eq.
(\ref{eq4}) tells us that the subtle distinction between space and
time of $S$ is that, in the expression of the invariant $\Delta
s^{2}$, the spatial coordinate $x$ is a real number and the temporal
term $ict$ is a purely imaginary number. This character can be
extracted as a criterion to distinguish between the concepts of
space and time.

Now substituting Eq. (\ref{eq3}) into Eq. (\ref{eq4}), we obtain the expression of $\Delta
s^{2}$ in $S^{\prime }$:

\begin{eqnarray}
\Delta s^{2}&=&\lbrack -i\frac{1}{\sqrt{\left( v/c\right) ^{2}-1}}(x-\frac{v}{c%
}ct)\rbrack ^{2}  \nonumber \\
&+&\lbrack -\frac{1}{\sqrt{\left( v/c\right) ^{2}-1}}(ct-\frac{%
v}{c}x)\rbrack ^{2}.\label{eq5}
\end{eqnarray}
Considering our criterion, we redefine the concepts of space and time in $%
S^{\prime }$:

\begin{equation}
\left\{
\begin{array}{l}
c\widetilde{t^{\prime }}=-\frac{1}{\sqrt{\left( v/c\right) ^{2}-1}}(x-\frac{v%
}{c}ct)\qquad (\rm{i}) \\
\widetilde{x^{\prime }}=-\frac{1}{\sqrt{\left( v/c\right) ^{2}-1}}(ct-\frac{v%
}{c}x)\quad \ \ \ \ (\rm{ii})
\end{array}
\right. .\label{eq6}
\end{equation}
Need to point out that we do not write out the expressions $\widetilde{%
y^{\prime }}=y,\widetilde{z^{\prime }}=z$ in Eq. (\ref{eq6}). Eq. (\ref{eq6}) implies

\begin{equation}
\left\{
\begin{array}{l}
\widetilde{x^{\prime }}-\rm{axis}:c\widetilde{t^{\prime }}%
=0\Longrightarrow x=\frac{v}{c}ct \\
c\widetilde{t^{\prime }}-\rm{axis}:\widetilde{x^{\prime }}%
=0\Longrightarrow x=\frac{c}{v}ct
\end{array}
\right. .\label{eq7}
\end{equation}

Comparing Eq. (\ref{eq7}) with Eq. (\ref{eq2}), we observe that, in
the superluminal inertial frame, the temporal coordinate exchanges
with the spatial coordinate in the motion direction. This can be
understood by the following way: for every point $P$ in the
space-time, we can express it with $(x,y,z,ct)$ in the rest frame
$S$ and with four real numbers $\left\{ q_{1}^{\prime
},q_{2}^{\prime
},q_{3}^{\prime },q_{4}^{\prime }\right\} $ in some inertial frame $%
S^{\prime }$. We just know something about $\left\{ q_{1}^{\prime
},q_{2}^{\prime },q_{3}^{\prime },q_{4}^{\prime }\right\} $ is that three of
them are spatial coordinates and the other one is temporal coordinate. If $%
S^{\prime }$ and $S$\ are selected according to the rules indicated at the
beginning of this article, we can take $q_{3}^{\prime }$ and $q_{4}^{\prime
} $ as spatial coordinates and set $q_{3}^{\prime }=y,q_{4}^{\prime }=z$.
Furthermore, we know that $\left\{ q_{1}^{\prime },q_{2}^{\prime }\right\} $
form an invariant

\begin{equation}
(\alpha q_{1}^{\prime })^{2}+(\beta q_{2}^{\prime
})^{2}=x^{2}+(ict)^{2},\label{eq8}
\end{equation}
where $\alpha $ and $\beta $ are two complex numbers: one is a real
number and another a purely imaginary number. If $\alpha $ is a
purely imaginary
number, then $q_{1}^{\prime }$\ represents the temporal coordinate and $%
q_{2}^{\prime }$ the spatial coordinate in $S^{\prime }$, and vice
versa. Note that $q_{1}^{\prime },q_{2}^{\prime },q_{3}^{\prime }$
and $q_{4}^{\prime }$ are real numbers because we regard them as
measurable magnitudes. In a word, the Lorentz transformations
between $S$ and $S^{\prime }$ are Eq. (\ref{eq6}) if $S^{\prime }$
is a superluminal inertial frame; otherwise take Eq. (\ref{eq1}).

Now suppose there is another inertial frame $S^{\prime \prime }$ moving at
the speed of $u\,(u=\frac{c^{2}}{v}<c)$ along $x$-axis of $S$. The Lorentz
transformations between $S$ and $S^{\prime \prime }$ is:

\begin{equation}
\left\{
\begin{array}{l}
x^{\prime \prime }=\frac{1}{\sqrt{1-\left( u/c\right) ^{2}}}(x-\frac{u}{c}ct)
\\
ct^{\prime \prime }=\frac{1}{\sqrt{1-\left( u/c\right) ^{2}}}(ct-\frac{u}{c}%
x)
\end{array}
\right. .\label{eq9}
\end{equation}
Substituting $u=\frac{c^{2}}{v}<c$ into Eq. (\ref{eq9}), we obtain:

\begin{equation}
\left\{
\begin{array}{l}
x^{\prime \prime }=-\frac{1}{\sqrt{\left( v/c\right) ^{2}-1}}(ct-\frac{v}{c}%
x) \\
ct^{\prime \prime }=-\frac{1}{\sqrt{\left( v/c\right) ^{2}-1}}(x-\frac{v}{c}%
ct)
\end{array}
\right. .\label{eq10}
\end{equation}
Obviously, Eq. (\ref{eq10}) coincides with Eq. (\ref{eq6}), which means that the coordinate
transformations between $S$ and\ $S^{\prime }\,(v>c)$ are the same as the
Lorentz transformations between $S$ and\ $S^{\prime \prime }\,(u<c)$. As is
known, the causality is abided by in the inertial frame $S^{\prime \prime }$
since $u<c$. Therefore, it can not be violated in the inertial frame $S^{\prime
} $ when $v>c$ as long as we redefine the concepts of space and time in $%
S^{\prime }$ according to Eq. (\ref{eq6}).

Now let us turn to the superluminal dynamics.

Based on SR, we can construct another invariant $\Psi $:

\begin{equation}
\Psi =xp_{x}+(ict)(iE/c),\label{eq11}
\end{equation}
where $p_{x}$ is the $x$-direction momentum of an object observed in some
inertial frame and $E$ is the total energy. Here we still omit the
magnitudes involving $y$ and $z$. It is evidence that the position of $x$ is
equivalent to that of $p_{x}$ in the expression $\Psi $. The case is
available to $ct$ and $E/c$. Therefore, we need to redefine the concepts of
momentums and energy in $S^{\prime }$ if $v>c$. Imitating Eq. (\ref{eq6}), we define
\begin{equation}
\left\{
\begin{array}{l}
\widetilde{E^{\prime }}/c=-\frac{1}{\sqrt{\left( v/c\right) ^{2}-1}}(p_{x}-%
\frac{v}{c}\frac{E}{c})\qquad (\rm{i}) \\
\widetilde{p_{x}^{\prime }}=-\frac{1}{\sqrt{\left( v/c\right) ^{2}-1}}(\frac{%
E}{c}-\frac{v}{c}p_{x})\quad \ \ \ \ \ \ (\rm{ii})
\end{array}
\right. .\label{eq12}
\end{equation}

Assume there is an object moving at the speed of $v\,(v>c)$ along $x$-axis
of $S$. Fixing $S^{\prime }$ on this object, we know that $\widetilde{%
p_{x}^{\prime }}=0$, and denote $E_{0}=\widetilde{E^{\prime }}$ called the
rest energy. Eq. (\ref{eq12}) suggests
\begin{equation}
E_{v}=E_{0}/\sqrt{1-(c/v)^{2}},\label{eq13}
\end{equation}
where the symbol $E$ in Eq. (\ref{eq12}) has been replaced by $E_{v}$.

\begin{figure}[htp!]
\includegraphics[width=10cm]{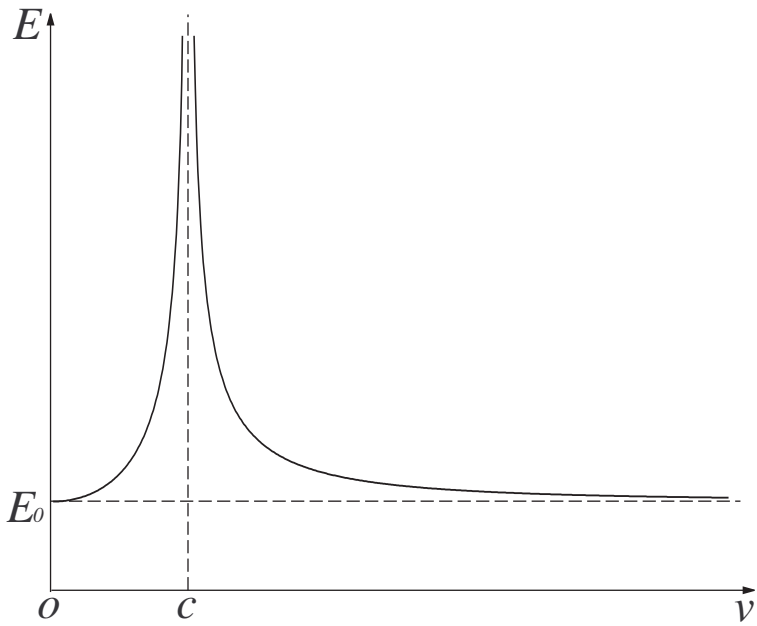}
\caption{The total energy $E$ of an object varying with the velocity
$v$. \label{fig2}}
\end{figure}

In fact, from SR we have known that the total energy and the rest energy has
the following relation
\begin{equation}
E_{u}=E_{0}/\sqrt{1-(u/c)^{2}}\label{eq14}
\end{equation}
when an object moves at the speed of $u\,(u<c)$ along $x$-axis of $S$. In
the same way, here we use $E_{u}$ to replace the symbol $E$.
Obviously, if $u=\frac{c^{2}}{v}$, then $E_{v}=E_{u}$, which means the total
energy of an object moving at a speed of $v$ $(v>c)$ is equal to the total
energy of that object moving at a speed of $u\,(u<c)$, and the product of
two speeds is $uv=c^{2}$. Considering SR and Eq. (\ref{eq13}), we can plot the curve
shown in Fig.\ref{fig2} to describe the total energy $E$ of an object varying with
the velocity $v$ (here $0<v<+\infty $ and $v\neq c$) if the rest energy $%
E_{0}>0$. As is shown in Fig.\ref{fig2}, the total energy $E$ increases with the
increasing of $v$ if $v<c$; and decreases with the increasing of $v$ if $v>c$%
. If $v\longrightarrow c$, then $E\longrightarrow +\infty $. As to the
mechanism that the object switches from the lower-than-$c$ state to
faster-than-$c$ state, it is a mystery.

The expression $uv=c^{2}$ is very familiar to us. Actually, we have
learned it in the de Broglie hypothesis of matter waves. This
hypothesis implies that the group velocity $u_{g}$ and the phase
velocity $v_{p}$ of matter waves satisfy the same expression
$u_{g}v_{p}=c^{2}$ \cite{k}. If this is not just a coincidence, the
further development our theory will give another novel
interpretation to the essence of matter waves.

The authors acknowledge Mr. I. Kanatchikov \cite{ik} and D. G.
Chakalov \cite{dg} who tell us that Mr. E. Recami {\it et
al.}\cite{er} proposed the similar ideas with us about 20 years ago
when our manuscript uploaded on arXiv.org.

\end{document}